\documentclass[12pt, nowcp]{jmlr}
\usepackage{atbegshi}
\usepackage{float}
\usepackage{lastpage}
\usepackage{emptypage}

\usepackage{listings}  
\usepackage{enumitem}

\lstnewenvironment{Code}
{
    \lstset{
        basicstyle=\ttfamily,          
        keepspaces=true,               
        columns=flexible,              
        breaklines=true,               
        commentstyle=\color{gray},     
        showstringspaces=false,        
        frame=single,                  
        framesep=3pt,                  
        framexleftmargin=1em,         
        xleftmargin=1em,              
        language=R                     
    }
}
{}

\newcommand{\code}[1]{\texttt{#1}}

\newcommand{\proglang}[1]{\textsf{#1}}

\newcommand{\pkg}[1]{\texttt{#1}}

\usepackage[utf8]{inputenc} 
\usepackage[T1]{fontenc}    
\usepackage{url}            
\usepackage{booktabs}       
\usepackage{amsfonts}       
\usepackage{nicefrac}       
\usepackage[protrusion=true, expansion=true]{microtype}
\usepackage{xcolor}         
\usepackage[format=plain,
            labelfont={bf,it},
            textfont=it]{caption}

\newcommand{\FRRMaxSpeedupGPUvsBaselineOverall}{42}
\newcommand{\FRRBaseRTimeNhundredDhundredDrawsTwoEFiveSec}{112}
\newcommand{\FRRGPUTimeNhundredDhundredDrawsTwoEFiveSec}{5}
\newcommand{\FRRSpeedupGPUvsBaselineNhundredDhundredDrawsTwoEFive}{24}
\newcommand{\FRRGPUVsCPUTimeReductionDthousandPct}{96}
\newcommand{\FRRCPUTimeNthousandDthousandDrawsTwoEFiveSec}{91}
\newcommand{\FRRGPUTimeNthousandDthousandDrawsTwoEFiveSec}{7}

\newcommand{\FRRvOnePctDThirty}{0.45}
\newcommand{\FRRvOnePctDHundred}{0.66}
\newcommand{\FRRvOnePctDThousand}{0.88}
\newcommand{\FRRRMSEDropPctDThirtyQOnePctRtwoFive}{15}
\newcommand{\FRRRMSEDropPctDThousandQOnePctRtwoFive}{3}
\newcommand{\FRRAcceptedQOnePctDrawsOneEFive}{1000}
\newcommand{\FRRMinPvalQOnePctDrawsOneEFive}{0.0010}
\newcommand{\FRRAcceptedQOnePctDrawsTwoEFive}{2000}
\newcommand{\FRRMinPvalQOnePctDrawsTwoEFive}{0.0005}

\newcommand{\FRRQStarDThousandRtwoFourTauPtTwo}{0.000012}
\newcommand{\FRRExpectedDrawsPerAcceptQStar}{81226}

\usepackage{placeins}

\usepackage{booktabs}
\usepackage{graphicx}
\usepackage{float}
\usepackage{lmodern}
\usepackage{latexsym,multirow}
\usepackage{amssymb,amsmath,bm}
\usepackage{bigints}
\usepackage{longtable}
\usepackage{bbm}
\usepackage[color,all,import,arrow]{xy}
\usepackage{array}
\usepackage{verbatim}
\usepackage{mathtools}
\usepackage{multicol}
\usepackage[utf8]{inputenc}
\usepackage{authblk}

\usepackage{dcolumn}
\newcolumntype{.}{D{.}{.}{-1}}
\newcolumntype{d}[1]{D{.}{.}{#1}}

\graphicspath{{./}}

\usepackage[T1]{fontenc}
\usepackage{tikz}
\usepackage{qtree}
\usepackage{comment}
\usetikzlibrary{arrows}
\usetikzlibrary{positioning}
\usetikzlibrary{matrix}
\usetikzlibrary{bayesnet}
\usetikzlibrary{decorations.pathreplacing,calligraphy}

\theoremheaderfont{\scshape}

\usepackage{rotating}

\usepackage{arydshln}

\usepackage[compact]{titlesec}

\usepackage{setspace}
\usepackage[bottom]{footmisc}
\allowdisplaybreaks

\makeatletter
\def\ps@jmlrtps{%
  \let\@mkboth\@gobbletwo
  \def\@oddhead{}%
  \def\@evenhead{}%
  \def\@oddfoot{}%
  \def\@evenfoot{}%
}
\makeatother

\title{\texttt{FastRerandomize}: Fast Rerandomization \\[0.4cm] Using Accelerated Computing}

\author[1]{Rebecca Goldstein}
\author[2]{Connor T. Jerzak}
\author[3]{Aniket Kamat}
\author[4]{Fucheng Warren Zhu}

\affil[1]{
   Robert F. Wagner School of Public Service, New York University, 
    \texttt{RebeccaSGoldstein.com} \authorcr
    \vspace{0.2cm}
}

\affil[2]{
    Department of Government, University of Texas at Austin,
    \texttt{ConnorJerzak.com} \authorcr
    \vspace{0.2cm}
}

\affil[3]{
    Department of Statistics, UC Berkeley, 
    \texttt{GitHub.com/aniketkamat} \authorcr
    \vspace{0.2cm}
}

\affil[4]{
    Departments of Statistics and Computer Science, Harvard University, \authorcr 
    \texttt{WarrenZhu.com} \authorcr
}

\date{}

\begin{document}

\maketitle

\vspace{-1.5cm}

\begin{abstract}%
  \noindent We present \pkg{fastrerandomize}, an \proglang{R} package for fast, scalable rerandomization in experimental design. Rerandomization improves precision by discarding treatment assignments that fail a prespecified covariate-balance criterion, but existing implementations can become computationally prohibitive as the number of units or covariates grows. \pkg{fastrerandomize} introduces three complementary advances: (i) optional GPU/TPU acceleration to parallelize balance checks, (ii) memory-efficient key-only storage that avoids retaining full assignment matrices, and (iii) auto-vectorized, just-in-time compiled kernels for batched candidate generation and inference. This approach enables exact or Monte Carlo rerandomization at previously intractable scales, making it practical to adopt the tighter balance thresholds required in modern high-dimensional experiments while simultaneously quantifying the resulting gains in precision and power for a given covariate set. Our approach also supports randomization-based testing conditioned on acceptance. In controlled benchmarks, we observe order-of-magnitude speedups over baseline workflows, with larger gains as the sample size or dimensionality grows, translating into improved precision of causal estimates. Code: \url{github.com/cjerzak/fastrerandomize-software}. Interactive capsule: \url{fastrerandomize.github.io/space}.
\end{abstract}

\vspace{0.5cm}
{
\small 
\noindent \textit{Author Note:} Forthcoming at \textit{SoftwareX}. Authors are listed in alphabetical order. We thank Kaz Sakamoto for feedback. The authors report no competing interests. 
}

\newpage 

\clearpage 

\section{Motivation and significance}
\label{sec:motivation}

\noindent Randomized experiments remain a cornerstone of empirical research across the social, biomedical, and other data-intensive sciences. While randomization guarantees unbiasedness, it can---especially in small samples or high-dimensional settings---produce treatment and control groups with appreciable covariate imbalance, inflating variance and reducing statistical power. Rerandomization addresses this challenge by repeatedly drawing candidate assignments and accepting only those that pass a prespecified balance criterion \citep{MorganRubin2012,li2018asymptotic}. Conditioning subsequent inference on the accepted set maintains valid design-based guarantees while improving precision.

Despite its conceptual simplicity, rerandomization can be computationally demanding. As the number of units grows, the space of possible assignments explodes combinatorially, and stringent balance thresholds may imply very low acceptance probabilities. In contemporary applications, covariate sets often include hundreds or thousands of features from text, images, or networks, exacerbating both runtime and memory pressure \citep{keith2020text,jerzak2023image,ogburn2024causal}. Practitioners might either forgo rerandomization or rely on ad hoc workflows that are difficult to scale.

Open-source software has advanced the state of practice in random assignment and randomization-based inference (e.g., \pkg{RItools}, \pkg{ri2}, \pkg{randomizeR}, \pkg{RATest}, \pkg{RCT2} \citep{RItools,ri2package,randomizeR,RATest,RCT2}). A smaller ecosystem targets rerandomization directly, including tools for power analysis, threshold selection, or iterative assignment generation \citep{branson2024power,kapelner2022optimal,jumble2024,NBERw23867}. Yet two persistent gaps remain: (i) scalable generation and storage of large pools of candidate randomizations under tight balance thresholds, and (ii) efficient, design-respecting inference once those pools are constructed.

\pkg{fastrerandomize} addresses these gaps through three complementary contributions:
\begin{enumerate}[leftmargin=*,nosep]
\item \textbf{Accelerated balance checks.} Batched, auto-vectorized kernels evaluate balance criteria across many candidates in parallel; GPU/TPU support further reduces latency.
\item \textbf{Key-only storage.} Instead of caching full assignment matrices, the backend retains compact pseudo-random keys sufficient to regenerate any accepted assignment on demand, greatly reducing memory requirements.
\item \textbf{Integrated design-based inference.} Exact or Monte Carlo generation integrates with randomization tests conditioned on acceptance, including optional fiducial intervals \citep{lehmann2005testing,leemis2020mathematical}.
\end{enumerate}

Together, these features extend rerandomization to experimental scales---large $n$, high-dimensional covariates, and tight balance thresholds---that were previously impractical, while maintaining the rigorous design-based guarantees emphasized in the theoretical literature \citep{MorganRubin2012,li2018asymptotic}. The practical need for such acceleration is most acute in modern applications where covariates are themselves outputs of deep learning pipelines---text embeddings, satellite-image descriptors, or neural-network features---that can yield hundreds or thousands of covariates per unit \citep{keith2020text,jerzak2023image,ogburn2024causal}. Achieving the level of balance needed to markedly improve precision and reduce $p$-hacking in high-dimensional, representation-rich experiments \citep{lu2025rerandomization} requires much more selective rerandomization rules and correspondingly larger pools of candidate assignments. \pkg{fastrerandomize} is designed to make this high-dimensional regime practically accessible: its accelerated balance checks and key-only storage allow researchers to adopt stringent, theoretically motivated thresholds in settings where covariates are derived from contemporary machine-learning models, without abandoning the rerandomization framework or incurring prohibitive computational cost.

\section{Software description}
\label{sec:software}

\subsection{Architecture}
\label{subsec:architecture}

\begin{figure}[htb]
\centering
\includegraphics[width=0.95\linewidth]{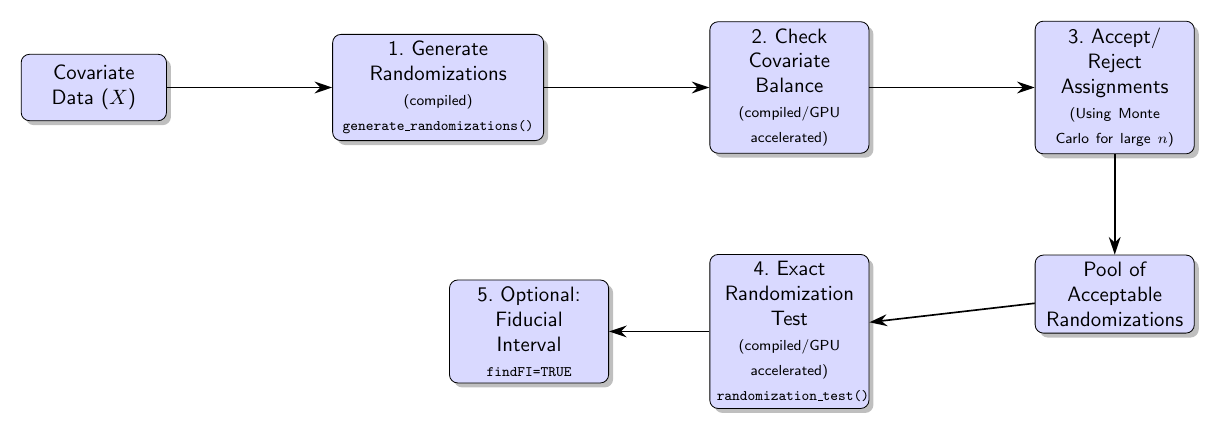}
\caption{
Core workflow in \pkg{fastrerandomize}. 
}
\label{fig:workflow}
\end{figure}

\pkg{fastrerandomize} follows a two-layer design (Figure~\ref{fig:workflow}):
\begin{itemize}[leftmargin=*,nosep]
\item An \proglang{R} front-end provides a compact API for assignment generation and inference, handling data validation, user-facing options, and I/O.
\item A JAX backend (managed via \pkg{reticulate}) compiles batched kernels for candidate generation, balance evaluation, and randomization tests. Where available, XLA (the Accelerated Linear Algebra compiler) compiles and dispatches these kernels to GPUs/TPUs; otherwise, optimized CPU kernels are used.
\end{itemize}

Two additional design choices underpin the scalability of randomization to previously computationally prohibitive scales:

\paragraph{Batched processing}
Candidate assignments are generated and evaluated in batches to limit peak memory use. Batching amortizes kernel launch and data transfer overhead while allowing auto-vectorized linear algebra to saturate available compute units.

\begin{figure}[htb]
\centering
\includegraphics[width=0.99\linewidth]{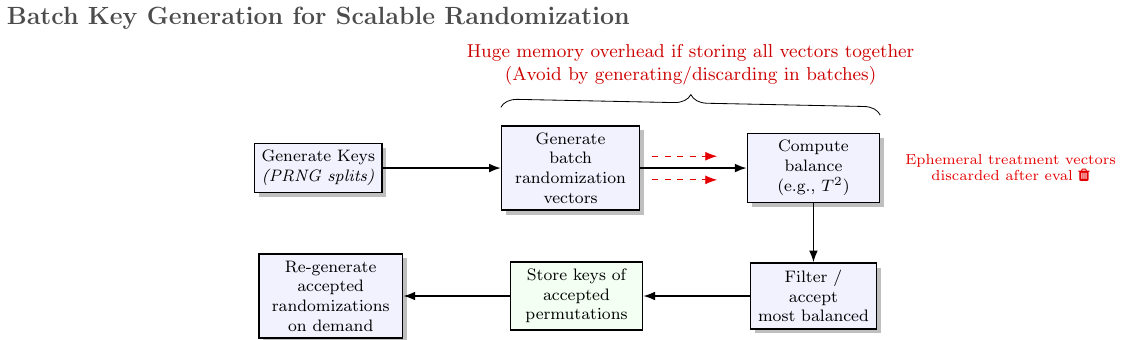}
\caption{
Key-only storage. Ephemeral assignment vectors are generated, checked for balance, and discarded; only keys for accepted draws are retained. Assignments can then be regenerated on demand for analysis, reducing memory by $\approx n/L$.
}
\label{fig:keys}
\end{figure}

\paragraph{Key-only storage}
For Monte Carlo workflows, \pkg{fastrerandomize} stores pseudorandom number generator (PRNG) \emph{keys} for accepted assignments (rather than full $n$-length vectors). When an assignment is needed---for plotting, export, or inference---the PRNG deterministically regenerates it from its key. If the key length is $L\ll n$, memory decreases by a factor of approximately $n/L$, enabling very large pools without overwhelming memory (Figure~\ref{fig:keys}). 

\subsection{Choosing stringency}
\label{subsec:diagnostic-theorem}

\noindent Having described the batched, accelerator-aware pipeline, we now turn to the practical question that governs how the software is used: \emph{given an observed imbalance distance, what does it imply for precision, and how stringent should rerandomization be?} 

We begin with a simple working model, drawing on foundations of \citep{MorganRubin2012,Li9157}. Let $X\in\mathbb{R}^{n\times d}$ denote pre-treatment covariates that have been column-standardized and whitened so that $\Sigma_X \approx I_d$, and (by a multivariate CLT / Gaussian approximation for sample mean differences) $\Delta_X$ is approximately $\mathcal{N}\!\left(0,\left(\tfrac{1}{n_T}+\tfrac{1}{n_C}\right) I_d\right)$. For a realized assignment, the treated–control mean difference is $\Delta_X=\bar X_T-\bar X_C$, and a resulting multivariate imbalance measure is
\[
M \;=\; \Delta_X^\top\Delta_X \;=\; \sum_{j=1}^d \mathrm{SMD}_j^2,
\]
reducing to the sum of squared \underline{s}tandardized \underline{m}ean \underline{d}ifferences (SMDs). A larger $M$ indicates greater imbalance; the goal is to understand how much that matters for precision and power.

Under a linear outcome model $Y_i(t)=\beta^\top X_i+\tau\,t+\varepsilon_i$ with $\operatorname{Var}(\varepsilon_i)=\sigma^2$, define the share of outcome variability attributable to covariates:
\[
R^2 \;\equiv\; \frac{\sigma_{\text{Prog}}^2}{\sigma_{\text{Prog}}^2+\sigma^2},
\qquad
\sigma_{\text{Prog}}^2
\equiv
\operatorname{Var}(\beta^\top X_i).
\]
A rotation-invariance argument (\ref{app:theorem}) implies the following \emph{typical-orientation} approximation for the difference-in-means RMSE targeting $\tau$:
\[
\operatorname{RMSE}_{\hat{\tau}}
\;\approx\;
\sqrt{\,
\sigma^2\!\left(\frac{1}{n_T}+\frac{1}{n_C}\right)
\;+\;
\frac{R^2}{1-R^2}\;\frac{\sigma^2}{d}\;M
\,},
\]
where $n_T$ and $n_C$ denote the treated and control group sizes, respectively. Intuitively, the first term is irreducible residual variance; the second term is a data-dependent penalty from the observed imbalance $M$. The power-relevant signal-to-noise ratio $|\tau|/\operatorname{RMSE}_{\hat{\tau}}$ therefore improves either by larger effects $|\tau|$ or by smaller $M$; once this ratio is already large enough for the study’s power target, tightening balance further will not change conclusions in a meaningful way.

Instead of setting a threshold on $M$, it is operationally simpler to specify an \emph{acceptance probability} $q$ (e.g., retaining the best $1\%$ of candidates). Since $M \stackrel{d}{\approx} (\tfrac{1}{n_T}+\tfrac{1}{n_C})\chi^2_d$ for whitened covariates, this strategy effectively truncates this $\chi^2_d$ distribution. Concretely, if $q\equiv \Pr(\text{accept})=\Pr(M\le a)$, then under the $\chi^2$ approximation we have
$q=\Pr(\chi^2_d\le c_q)$ with $c_q \equiv F^{-1}_{\chi^2_d}(q)$, and the implied threshold on $M$ is
$a_q = (\frac{1}{n_T}+\frac{1}{n_C})\,c_q$.
Under this rule, the shrinkage factor can be written as
$v_a(d)=\mathbb{E}[\chi^2_d\mid \chi^2_d\le c_q]/d = \Pr(\chi^2_{d+2}\le c_q)/q$
(see \ref{app:theorem}). Geometrically, this restricts $\Delta_X$ to a tighter hypersphere, shrinking the imbalance covariance by a scalar factor $v_a(d) \in (0,1)$, which we compute in \code{diagnose\_rerandomization()}. This shrinkage directly reduces the expected MSE:
\[
\mathbb{E}\!\left[\mathrm{MSE}(\widehat\tau_{\mathrm{DiM}})\,\middle|\,\text{accept}\right]
\;=\;
\left(\tfrac{1}{n_T}+\tfrac{1}{n_C}\right)\Big(\sigma^2 + v_a(d)\,\sigma_{\text{Prog}}^2\Big),
\quad
\text{where}\;\; \sigma_{\text{Prog}}^2=\tfrac{R^2}{1-R^2}\,\sigma^2.
\]
Thus, increasing stringency (lowering $q$) leaves irreducible residual variance ($\sigma^2$) unchanged while systematically attenuating the error contribution from prognostic imbalance.

\begin{figure}[htb]
\centering
\includegraphics[width=0.85\linewidth]{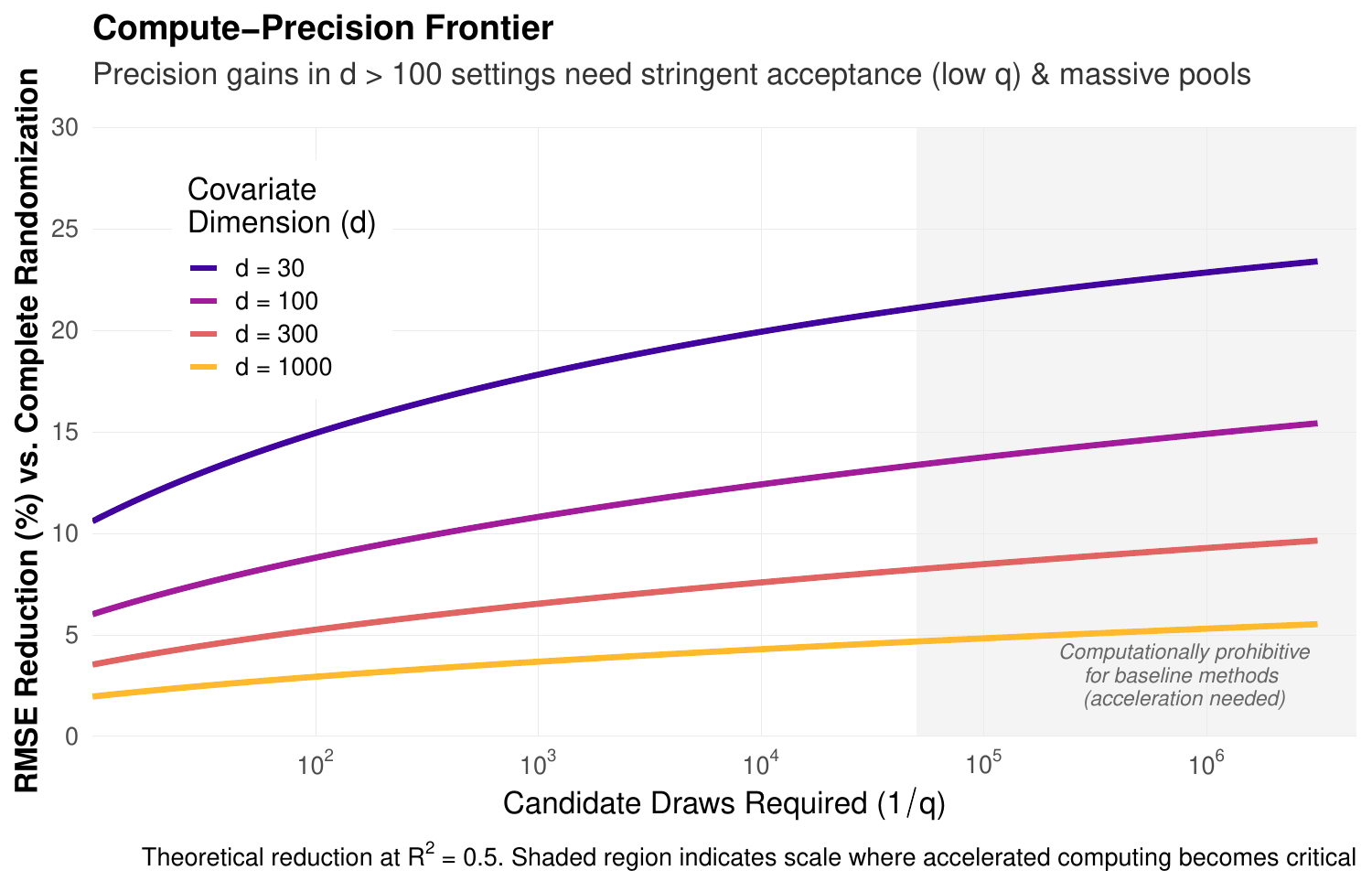}
\caption{
\textbf{The Compute-Precision Frontier.} Theoretical RMSE reduction (\%) versus complete randomization (assuming $R^2=0.5$) as a function of required candidate draws ($1/q$) across dimensions $d \in \{30, 100, 300, 1000\}$. Achieving meaningful precision gains in high-dimensional settings ($d > 100$) requires stringent acceptance criteria.}
\label{fig:RMSEModel}
\end{figure}

As a rule of thumb, at $q{=}0.01$ acceptance, the expected scaling term on the covariate-imbalance factor is $v_a(d)= \FRRvOnePctDThirty{}$ at $d{=}30$, it is $\FRRvOnePctDHundred{}$ at $d{=}100$, and it is $\FRRvOnePctDThousand{}$ at $d{=}1000$ (larger means greater contribution to MSE from imbalance). In precision terms, this translates, at $n_T{=}n_C{=}500$ and $R^2{=}0.5$, to an RMSE reduction of about $\FRRRMSEDropPctDThirtyQOnePctRtwoFive{}\%$ at $d{=}30$, but only $\FRRRMSEDropPctDThousandQOnePctRtwoFive{}\%$ at $d{=}1000$. Therefore, in very high-$d$ settings, even more selective acceptance rates are necessary to yield the same precision gains as in small dimensionality settings \citep{MorganRubin2012}---further motivating the need for the accelerated computing approach taken here. Figure~\ref{fig:RMSEModel} illustrates this compute-precision frontier, showing that while modest pools suffice for low dimensions, achieving comparable precision gains in high-dimensional settings requires stringent acceptance rates (low $q$) and therefore larger candidate pools. 

Two recurring planning questions follow naturally. First, \emph{does drawing more candidates improve balance?} No: balance is governed by $q$, increasing draws mainly increases the number of accepted assignments and hence the Monte Carlo resolution. Holding $q{=}0.01$ fixed, going from $10^5$ to $2\times 10^5$ candidates simply doubles the accepted set (from $\FRRAcceptedQOnePctDrawsOneEFive{}$ to $\FRRAcceptedQOnePctDrawsTwoEFive{}$ assignments) and halves the minimum attainable randomization-test $p$ (from $\FRRMinPvalQOnePctDrawsOneEFive{}$ to $\FRRMinPvalQOnePctDrawsTwoEFive{}$); balance itself is unchanged.

Second, \emph{how should $q$ be chosen?} A simple planning heuristic is that rerandomization at level $q$ typically needs about $1/q$ candidate draws per accepted assignment. A power-focused approach simplifies $q$ selection: specify $(n_T,n_C)$, an outcome scale $\sigma$, a plausible $R^2$, and the effect size of interest $|\tau|$ with desired power and size. In an example high-dimensional case ($d{=}1000$, $n_T{=}n_C{=}500$, $\sigma{=}1$, $R^2{=}0.4$), targeting a two-sided $\alpha{=}0.05$ test with $80\%$ power for $|\tau|{=}0.2$ implies a recommended acceptance probability of $q^\star \approx \FRRQStarDThousandRtwoFourTauPtTwo{}$, or about $\FRRExpectedDrawsPerAcceptQStar{}$ draws per accepted assignment. This inversion is implemented by \code{diagnose\_rerandomization()}.

In sum, we recommend the following iteration cycle for selecting imbalance thresholds in a given scenario: 
\begin{enumerate}[leftmargin=*,nosep]
\item Start with a coarse $q$, compute $\operatorname{RMSE}_{\hat{\tau}}$ from the observed $M$, using parameters from prior literature.
\item Tighten $q$ if $|\tau|/\operatorname{RMSE}$ is below the power target.
\item Stop when additional tightening results in minimal change.
\end{enumerate}

\subsection{Software functionalities}
\label{subsec:functionalities}

\noindent The package exposes four core functions. Further details are provided in Appendix B.

\paragraph{\code{build\_backend()}}
Creates a minimal conda environment (e.g., named \code{"fastrerandomize"}) with Python and JAX dependencies. If a GPU or TPU device is present, the package selects an appropriate accelerated backend.

\paragraph{\code{generate\_randomizations()}}
Constructs pools of acceptable assignments under a specified balance criterion:
\begin{itemize}[leftmargin=*,nosep]
\item \textbf{Exact enumeration} (\code{randomization\_type = "exact"}): systematically iterates over all possible assignments when feasible.
\item \textbf{Monte Carlo} (\code{"monte\_carlo"}): draws and filters batched candidates, storing keys for accepted draws.
\end{itemize}
Users control stringency via \code{randomization\_accept\_prob}. Hotelling's $T^2$ is the default balance metric; a custom vectorized function can be supplied.

\paragraph{\code{diagnose\_rerandomization()}}
Maps the diagnostics in Section~\ref{subsec:diagnostic-theorem} into a single planning tool. Given either an observed imbalance summary or design-stage inputs $(n_T,n_C,d,R^2,\sigma,|\tau|)$, the function reports (i) the implied realized RMSE and a conservative upper bound, (ii) the expected ex-ante RMSE under a given acceptance probability $q$, and (iii) the largest $q$ that meets a user-supplied precision or power target. 

\paragraph{\code{randomization\_test()}}
Implements design-respecting inference conditional on acceptance: the observed statistic (e.g., difference-in-means) is compared to its distribution across accepted assignments, yielding an exact or approximate $p$-value for the sharp null \citep{lehmann2005testing,MorganRubin2012,li2018asymptotic} (depending on whether full enumeration or Monte Carlo is used). Optional inversion returns a fiducial interval \citep{leemis2020mathematical}. Because only accepted assignments enter the reference distribution, $p$-values are lower-bounded by $1/M_{\text{Accept}}$, where $M_{\text{Accept}}$ is the number of accepted draws \citep{lehmann2005testing}.

\section{Illustrative example and workflow}
\label{sec:examples}

\noindent High-dimensional pre-treatment features---e.g., text embeddings, satellite-image descriptors, or network-derived covariates---are increasingly common \citep{keith2020text,jerzak2023image,ogburn2024causal}. \pkg{fastrerandomize} is designed to enable stringent rerandomization in these settings while preserving design-based guarantees.

To fix ideas, consider a field experiment that randomizes roughly $n \approx 1000$ villages into treatment and control. For each village, investigators construct a high-dimensional vector $X_i$ from strictly pre-treatment satellite imagery, using a CLIP-based neural network image embedding model to extract the 768 features summarizing settlement structure, roads, vegetation, and other context \citep{xiao2025foundation,jerzak2023image}. Researchers would like villages to be well-matched on these remotely sensed covariates, so that outcomes are not confounded by local conditions visible from above. Achieving this with a tight rerandomization rule---for example, keeping only the best $1\%$ of candidate assignments by imbalance in $X$---can substantially shrink the imbalance component of the MSE (\ref{subsec:diagnostic-theorem}), but requires generating on the order of $1/q$ candidate draws per accepted assignment. With $d$ in the hundreds, such regimes quickly become intractable for existing methods. A typical \pkg{fastrerandomize} workflow in this context is:

\begin{enumerate}[leftmargin=*,nosep]
\item \textbf{Prepare covariates.} Assemble the $n\times d$ matrix $X$ (rows are units, columns are features). Column-wise standardization is recommended. Covariates predictive of outcome should be prioritized for inclusion in $X$.

\item \textbf{Choose stringency.} Set \code{randomization\_accept\_prob} to the desired acceptance probability, $q$, following the diagnostic approach outlined in Section~\ref{subsec:diagnostic-theorem} in order to model balance-compute-power trade-offs. Holding $q$ fixed, increasing the number of draws enlarges the accepted pool (lowering the minimum attainable randomization-test $p$-value) without changing expected balance. \code{diagnose\_rerandomization()} can be used to select an acceptance probability or threshold consistent with a target precision/power.

\item \textbf{Generate candidates at scale.} Call \code{generate\_randomizations()} (usually with \code{randomization\_type = "monte\_carlo"}), using batched draws and, where available, GPU/TPU acceleration to evaluate balance statistics in parallel. Internally, \pkg{fastrerandomize} relies on key-only storage to avoid holding large binary assignment arrays in memory.
\item \textbf{Infer with design respect.} Use \code{randomization\_test()} to compute $p$-values (and optionally fiducial intervals) from the distribution of the test statistic over the \emph{accepted} assignments, preserving design-based validity under either exact or Monte Carlo generation.
\end{enumerate}

\section{Impact}
\label{sec:impact}

\begin{figure}[htb]
\centering
\includegraphics[width=1.1\linewidth]{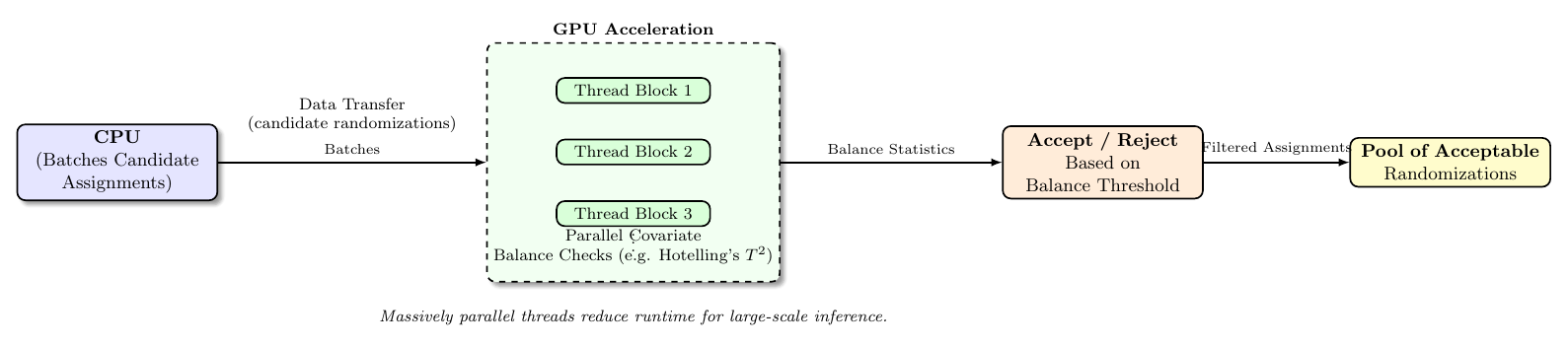}
\caption{GPU acceleration: CPU forms batched candidates; balance checks are parallelized across many compute units on the GPU.}
\label{fig:GPUExplain}
\end{figure}

\paragraph{Enabling stringent balance at scale}
Rerandomization improves the precision of treatment effect estimators by conditioning on good covariate balance \citep{MorganRubin2012,li2018asymptotic}. In practice, stringent thresholds can sharply reduce acceptance probabilities, requiring large candidate pools. \pkg{fastrerandomize} makes this regime tractable through batched, compiled kernels and optional GPU/TPU acceleration (Figure~\ref{fig:GPUExplain}). In comparative simulations, we observe speedups over baseline implementations exceeding $\FRRMaxSpeedupGPUvsBaselineOverall{}\times$ in high-$d$ and high-$n$ regimes and when requiring more stringent acceptance thresholds. 

\begin{figure}[htb]
\centering
\includegraphics[width=0.95\linewidth]{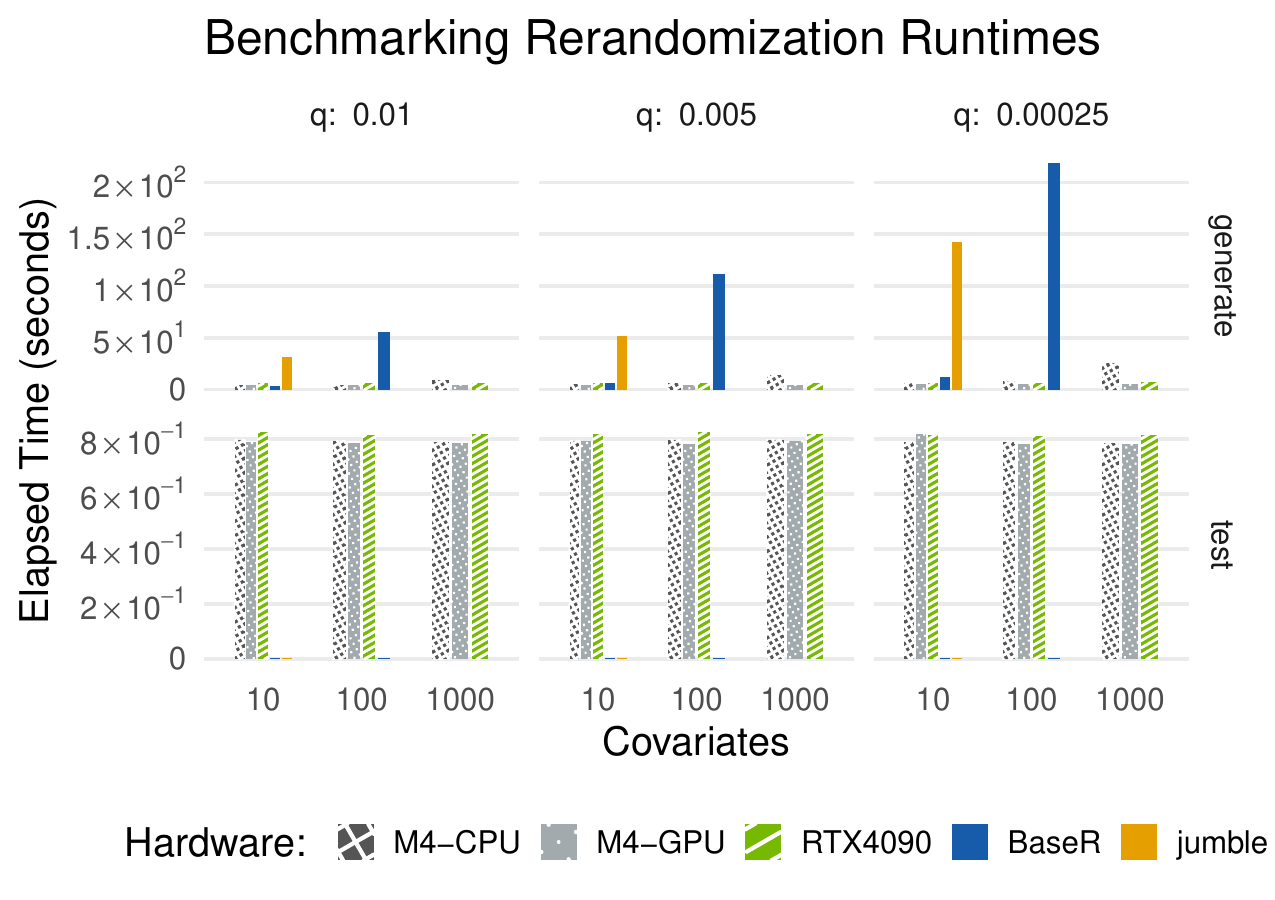}
\caption{
Baselines and \texttt{fastrerandomize}'s CPU/GPU runtimes for $n{=}100$. Bars show elapsed time (seconds) for (top) generating large pools of randomizations and (bottom) randomization-based inference, across covariate dimensions and rerandomization acceptance thresholds ($q \in \{$0.01, 0.005, 0.00025$\}$). Base R and \pkg{jumble} implementations are shown only for scenarios in which they complete within a reasonable time frame; in larger-scale settings or when $n\leq d$, they become computationally prohibitive or do not produce results with default settings.
} 
\label{fig:bench100}
\end{figure}

\begin{figure}[htb]
\centering
\includegraphics[width=0.95\linewidth]{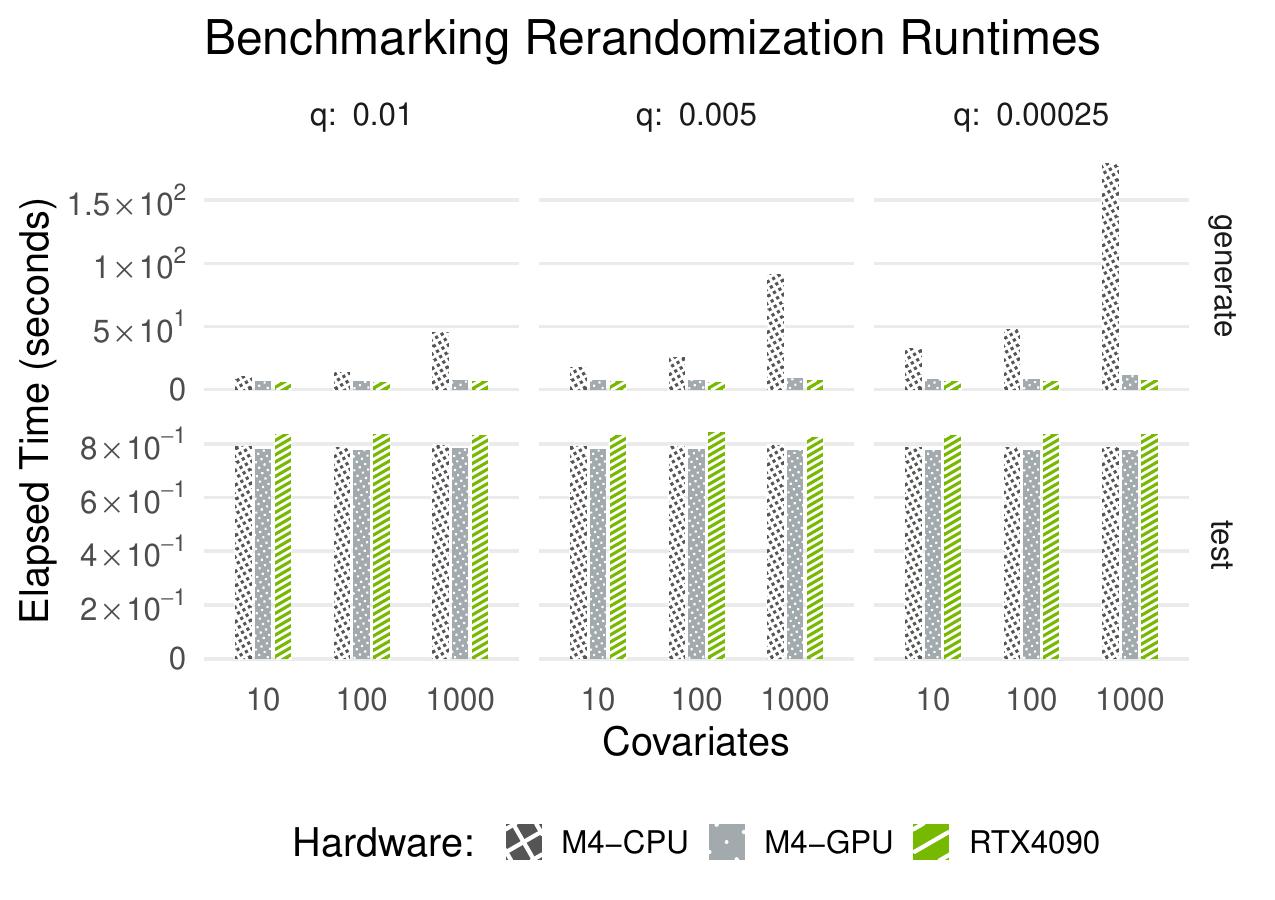}
\caption{\texttt{fastrerandomize} with CPU vs.\ GPU for $n{=}1000$. Bars show elapsed time (seconds) for (top) pool generation and (bottom) inference, across covariate dimensions and acceptance probability for rerandomization.
}
\label{fig:bench1000}
\end{figure}

\begin{figure}[htb]
\centering
\includegraphics[width=0.75\linewidth]{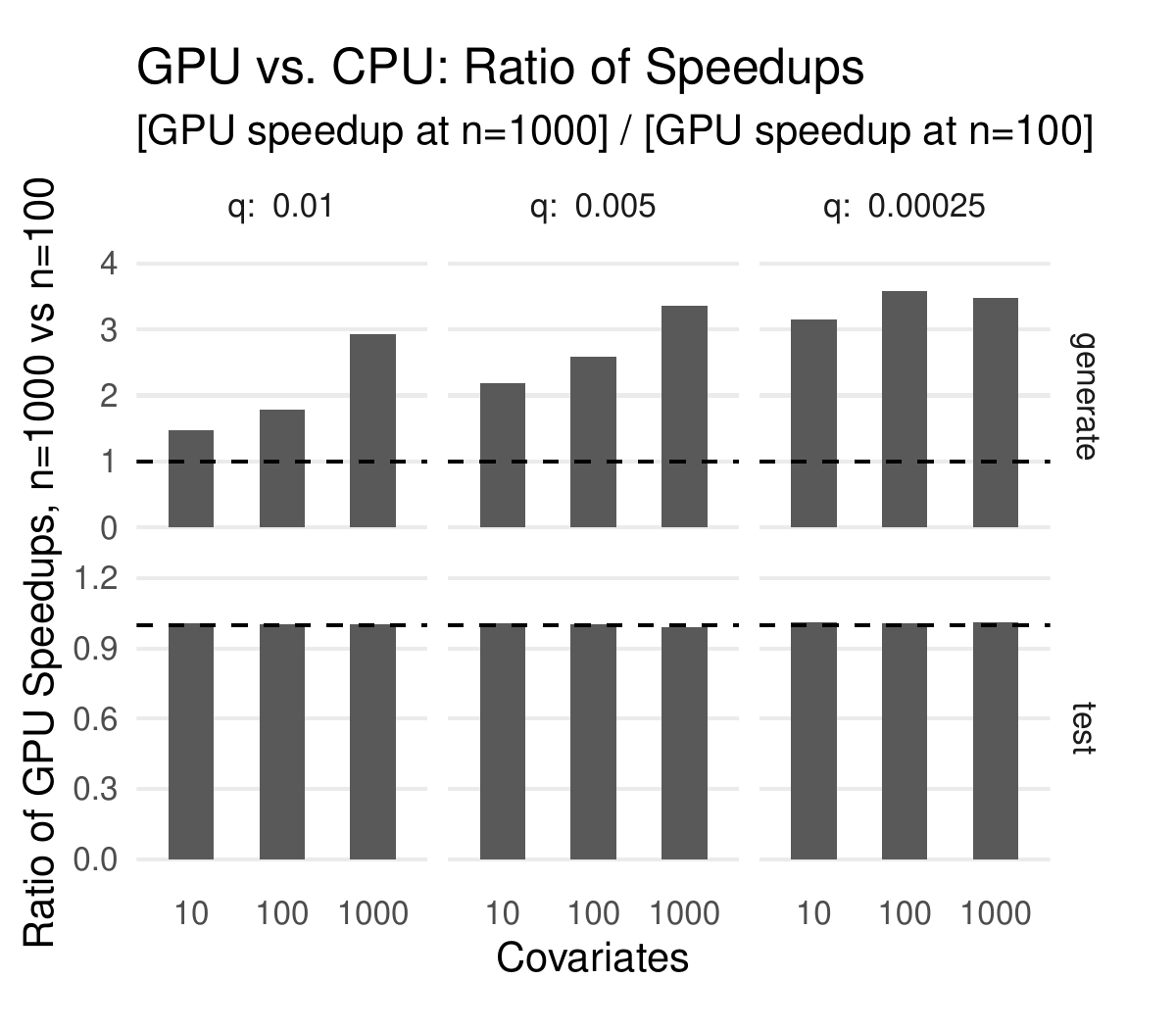}
\caption{
GPU advantage grows with problem size. Bars plot the ratio of GPU (M4) speedups at $n{=}1000$ relative to $n{=}100$ (values $>1$ indicate larger gains at $n{=}1000$), by draw count and step (pool generation vs.\ inference).
}
\label{fig:ratioSpeedups}
\end{figure}

To demonstrate, we report controlled benchmarks that vary sample size ($n$), covariate dimension ($d$), and Monte Carlo draw size (Figures~\ref{fig:bench100}-\ref{fig:bench1000}). For $n{=}100$ and $d{=}100$ with $q=0.005$, the GPU backend (both Apple M4 [METAL] and NVIDIA RTX 4090 [CUDA]) completes pool generation in about \FRRGPUTimeNhundredDhundredDrawsTwoEFiveSec{}\,s, versus \,\FRRBaseRTimeNhundredDhundredDrawsTwoEFiveSec{}\,s for a baseline \proglang{R} approach ($\approx$ \FRRSpeedupGPUvsBaselineNhundredDhundredDrawsTwoEFive{}$\times$ speedup). We also benchmarked the recent \pkg{jumble} package~\citep{jumble2024}, currently one of the only other \proglang{R} tools that perform rerandomization with a user-specified acceptance rate. \pkg{jumble}'s performance lags the pure-\proglang{R} baseline, confirming that our gains stem from batched, XLA-based compilation and hardware acceleration rather than minor \proglang{R} implementation differences. Relative to the XLA-optimized \pkg{fastrerandomize} CPU backend, GPU execution reduces runtime by up to \FRRGPUVsCPUTimeReductionDthousandPct{}\% at $d{=}1000$. At $n{=}1000$ and $d{=}1000$, pool generation falls from \FRRCPUTimeNthousandDthousandDrawsTwoEFiveSec{}\,s (CPU) to \,\FRRGPUTimeNthousandDthousandDrawsTwoEFiveSec{}\,s (GPU) at $q=0.005$. Figure \ref{fig:ratioSpeedups} shows how the GPU advantage increases with problem size in $n$.

\paragraph{Reducing memory pressure}
Key-only storage decouples \emph{evaluation} from \emph{retention}. Assignments are generated and evaluated in place; only the compact keys for accepted draws persist. If a key is two integers ($L{=}2$) and $n{=}10^3$, retaining keys instead of full $0/1$ vectors reduces memory by roughly a factor of $n/L \approx 500$ (Figure~\ref{fig:keys}). This enables consideration of stringent randomization thresholds without exhausting memory.

\paragraph{Design-based inference}
By returning a structured object for accepted draws, \pkg{fastrerandomize} integrates well with randomization tests conditioned on acceptance structure. $p$-values are computed by comparing the observed statistic to its distribution across \emph{accepted} assignments; inversion yields intervals \citep{lehmann2005testing,leemis2020mathematical}. 

\paragraph{Compatibility and ecosystem}
\pkg{fastrerandomize} complements established experimental design and inference packages \citep{RItools,ri2package,randomizeR,RATest,RCT2} by focusing on the scalable construction of \emph{balanced} assignment pools and on accelerated, design-respecting inference. For rerandomization-specific tasks such as power analysis and threshold tuning \citep{branson2024power,kapelner2022optimal,jumble2024,NBERw23867}, \pkg{fastrerandomize} can serve as a downstream engine for candidate generation and testing at scale. The package also includes \code{fast\_distance()} for accelerated pairwise distance calculations in applications beyond randomization, such as phylogenetics \citep{elias2007fast} and genomics \citep{akhtyamov2024gpu}. 

\section{Conclusions}
\label{sec:conclusions}

\noindent \pkg{fastrerandomize} brings rerandomization into the realm of large-scale, high-dimensional experimentation by combining batched computation, hardware acceleration, and memory-efficient storage with design-based inference. Practitioners can now adopt stringent balance thresholds---even with thousands of units and covariates---without prohibitive runtime or memory costs. Looking ahead, promising avenues include multi-device parallelism for candidate generation and richer balance criteria tailored to structured text, image, or network covariates \citep{keith2020text,jerzak2023image,ogburn2024causal}.

\clearpage\newpage 
\bibliographystyle{elsarticle-num}
\bibliography{mybib}

\begin{thebibliography}{23}
\providecommand{\natexlab}[1]{#1}
\providecommand{\url}[1]{\texttt{#1}}
\expandafter\ifx\csname urlstyle\endcsname\relax
  \providecommand{\doi}[1]{doi: #1}\else
  \providecommand{\doi}{doi: \begingroup \urlstyle{rm}\Url}\fi

\bibitem[Akhtyamov et~al.(2024)Akhtyamov, Nabi, Gafurov, Sizykh, Favorov,
  Medvedeva, and Stupnikov]{akhtyamov2024gpu}
Pavel Akhtyamov, Ausaaf Nabi, Vladislav Gafurov, Alexey Sizykh, Alexander
  Favorov, Yulia Medvedeva, and Alexey Stupnikov.
\newblock {GPU-accelerated Kendall Distance Computation for Large or Sparse
  Data}.
\newblock \emph{GigaScience}, 13:\penalty0 giae088, 2024.

\bibitem[Banerjee et~al.(2017)Banerjee, Chassang, Montero, and
  Snowberg]{NBERw23867}
Abhijit Banerjee, Sylvain Chassang, Sergio Montero, and Erik Snowberg.
\newblock {A Theory of Experimenters}.
\newblock Working Paper 23867, National Bureau of Economic Research, September
  2017.
\newblock URL \url{http://www.nber.org/papers/w23867}.

\bibitem[Bowers et~al.(2023)Bowers, Fredrickson, and Hansen]{RItools}
Jake Bowers, Mark Fredrickson, and Ben Hansen.
\newblock \emph{{RItools: Randomization Inference Tools}}, 2023.
\newblock URL \url{https://CRAN.R-project.org/package=RItools}.
\newblock {R package version 0.3-3}.

\bibitem[Branson et~al.(2024)Branson, Li, and Ding]{branson2024power}
Zach Branson, Xinran Li, and Peng Ding.
\newblock {Power and Sample Size Calculations for Rerandomization}.
\newblock \emph{Biometrika}, 111\penalty0 (1):\penalty0 355--363, 2024.

\bibitem[Coppock(2022)]{ri2package}
Alexander Coppock.
\newblock \emph{{ri2: Randomization Inference for Randomized Experiments}},
  2022.
\newblock URL \url{https://CRAN.R-project.org/package=ri2}.
\newblock {R package version 0.4.0}.

\bibitem[Coppock(2024)]{randomizeR}
Alexander Coppock.
\newblock \emph{{randomizr: Easy-to-Use Tools for Common Forms of Random
  Assignment and Sampling}}, 2024.
\newblock URL \url{https://github.com/DeclareDesign/randomizr}.
\newblock {R package version 1.0.0}.

\bibitem[Elias and Lagergren(2007)]{elias2007fast}
Isaac Elias and Jens Lagergren.
\newblock {Fast Computation of Distance Estimators}.
\newblock \emph{BMC Bioinformatics}, 8\penalty0 (1):\penalty0 89, 2007.

\bibitem[Huang et~al.(2022)Huang, Jiang, and Imai]{RCT2}
Karissa Huang, Zhichao Jiang, and Kosuke Imai.
\newblock \emph{Designing and Analyzing Two-Stage Randomized Experiments},
  October 16 2022.
\newblock URL \url{https://CRAN.R-project.org/package=RCT2}.
\newblock {R package version 0.0.1}.

\bibitem[Jerzak et~al.(2023)Jerzak, Johansson, and Daoud]{jerzak2023image}
Connor~T. Jerzak, Fredrik Johansson, and Adel Daoud.
\newblock {Image-based Treatment Effect Heterogeneity}.
\newblock \emph{Proceedings of the Second Conference on Causal Learning and
  Reasoning (CLeaR), Proceedings of Machine Learning Research (PMLR)},
  213:\penalty0 531--552, 2023.

\bibitem[Kapelner et~al.(2022)Kapelner, Krieger, Sklar, and
  Azriel]{kapelner2022optimal}
Adam Kapelner, Abba~M Krieger, Michael Sklar, and David Azriel.
\newblock {Optimal Rerandomization Designs via a Criterion That Provides
  Insurance Against Failed Experiments}.
\newblock \emph{Journal of Statistical Planning and Inference}, 219:\penalty0
  63--84, 2022.

\bibitem[Keith et~al.(2020)Keith, Jensen, and O’Connor]{keith2020text}
Katherine Keith, David Jensen, and Brendan O’Connor.
\newblock {Text and Causal Inference: A Review of Using Text to Remove
  Confounding from Causal Estimates}.
\newblock In \emph{Proceedings of the 58th Annual Meeting of the Association
  for Computational Linguistics}, pages 5332--5344, 2020.

\bibitem[Leemis(2020)]{leemis2020mathematical}
Lawrence Leemis.
\newblock \emph{Mathematical Statistics}.
\newblock Ascended Ideas, 2020.
\newblock ISBN 9780982917466.

\bibitem[Lehmann and Romano(2005)]{lehmann2005testing}
E.~L. Lehmann and Joseph~P. Romano.
\newblock \emph{{Testing Statistical Hypotheses}}.
\newblock Springer Texts in Statistics. Springer, New York, third edition,
  2005.
\newblock ISBN 0-387-98864-5.

\bibitem[Li et~al.(2018{\natexlab{a}})Li, Ding, and Rubin]{Li9157}
Xinran Li, Peng Ding, and Donald~B. Rubin.
\newblock {Asymptotic Theory of Rerandomization in Treatment-Control
  Experiments}.
\newblock \emph{Proceedings of the National Academy of Sciences}, 115\penalty0
  (37):\penalty0 9157--9162, 2018{\natexlab{a}}.
\newblock \doi{10.1073/pnas.1808191115}.
\newblock URL \url{https://www.pnas.org/content/115/37/9157}.

\bibitem[Li et~al.(2018{\natexlab{b}})Li, Ding, and Rubin]{li2018asymptotic}
Xinran Li, Peng Ding, and Donald~B Rubin.
\newblock {Asymptotic Theory of Rerandomization in Treatment--control
  Experiments}.
\newblock \emph{Proceedings of the National Academy of Sciences}, 115\penalty0
  (37):\penalty0 9157--9162, 2018{\natexlab{b}}.

\bibitem[Lu and Ding(2025)]{lu2025rerandomization}
Xin Lu and Peng Ding.
\newblock {Rerandomization for Covariate Balance Mitigates p-hacking in
  Regression Adjustment}.
\newblock \emph{arXiv preprint arXiv:2505.01137}, 2025.

\bibitem[McConeghy(2024)]{jumble2024}
Kevin McConeghy.
\newblock {jumble: An {R} Package to Perform Stratified and Re-randomization
  Procedures and Assess Covariate Balance}, 2024.
\newblock URL \url{https://github.com/kmcconeghy/jumble}.
\newblock {R package for clinical trial randomization}.

\bibitem[Morgan and Rubin(2012)]{MorganRubin2012}
Kari~Lock Morgan and Donald~B. Rubin.
\newblock {Rerandomization to Improve Covariate Balance in Experiments}.
\newblock \emph{The Annals of Statistics}, 40\penalty0 (2):\penalty0
  1263--1282, 2012.
\newblock ISSN 0090-5364, 2168-8966.
\newblock URL
  \url{http://projecteuclid.org.ezp-prod1.hul.harvard.edu/euclid.aos/1342625468}.

\bibitem[Ogburn et~al.(2024)Ogburn, Sofrygin, Diaz, and Van~der
  Laan]{ogburn2024causal}
Elizabeth~L Ogburn, Oleg Sofrygin, Ivan Diaz, and Mark~J Van~der Laan.
\newblock {Causal Inference for Social Network Data}.
\newblock \emph{Journal of the American Statistical Association}, 119\penalty0
  (545):\penalty0 597--611, 2024.

\bibitem[Olivares-Gonzalez and Sarmiento-Barbieri(2017)]{RATest}
Mauricio Olivares-Gonzalez and Ignacio Sarmiento-Barbieri.
\newblock \emph{Randomization Tests}, 2017.
\newblock URL \url{https://github.com/ignaciomsarmiento/RATest}.
\newblock {R package version 0.1.4}.

\bibitem[Rosner et~al.(2006)]{rosner2006fundamentals}
Bernard~A Rosner et~al.
\newblock \emph{{Fundamentals of Biostatistics}}, volume~6.
\newblock Thomson-Brooks/Cole Belmont, CA, 2006.

\bibitem[Vershynin(2025)]{vershynin2025high}
Roman Vershynin.
\newblock \emph{High-dimensional Probability: An Introduction with Applications
  in Data Science (Second Edition}.
\newblock Cambridge University Press, 2025.

\bibitem[Xiao et~al.(2025)Xiao, Xuan, Wang, Huang, Tao, Lu, and
  Yokoya]{xiao2025foundation}
Aoran Xiao, Weihao Xuan, Junjue Wang, Jiaxing Huang, Dacheng Tao, Shijian Lu,
  and Naoto Yokoya.
\newblock {Foundation Models for Remote Sensing and Earth Observation: A
  Survey}.
\newblock \emph{IEEE Geoscience and Remote Sensing Magazine}, 2025.

\end{thebibliography}

\singlespacing

\appendix
\section{Derivations: From observed distance to target MSE}
\label{app:theorem}

\noindent We herein justify the formal expressions found in the main text. Throughout, let $n_T+n_C=n$, and assume Bernoulli($\pi$) or complete randomization with fixed $n_T$; replace $n_T,n_C$ by $n\pi,n(1-\pi)$ when a design-stage approximation is desired (i.e., treating $n_T \approx n\pi$ and $n_C \approx n(1-\pi)$). We do not claim novelty in these analyses, but include them here for clarity of exposition; for more information, see the foundational work of \cite{MorganRubin2012,Li9157}. 

\subsection*{A. Conditional (realized) bias, variance, and MSE}

\noindent Write $\bar Z_T = n_T^{-1}\sum_{i:T_i=1} Z_i$ and $\bar Z_C$ analogously. Under the linear model, 
\[
Y_i(t) \;=\; \beta^\top X_i + \tau\, t + \varepsilon_i,\qquad \varepsilon_i \stackrel{\text{iid}}{\sim} \mathcal{N}(0,\sigma^2),
\]
the difference-in-means is
\[
\widehat\tau_{\mathrm{DiM}} \;=\; \bar Y_T - \bar Y_C
\;=\; \tau + \beta^\top\Delta_X + \big(\bar\varepsilon_T - \bar\varepsilon_C\big),
\quad
\Delta_X \equiv \bar X_T - \bar X_C.
\]
Conditioning on $(X,T)$, the noise term, assumed to be exogenous, has mean $0$ and variance $\sigma^2\!\left(\tfrac{1}{n_T}+\tfrac{1}{n_C}\right)$. Consequently, we can write the following conditional error/offset term due to imbalance and a variance term due to sampling: 
\begin{align*}
\mathrm{Bias}(\widehat\tau_{\mathrm{DiM}}\,|\,X,T) &= \beta^\top\Delta_X,\\
\mathrm{Var}(\widehat\tau_{\mathrm{DiM}}\,|\,X,T) &= \sigma^2\!\left(\tfrac{1}{n_T}+\tfrac{1}{n_C}\right),\\
\mathrm{MSE}(\widehat\tau_{\mathrm{DiM}}\,|\,X,T) &= \mathrm{Bias}(\widehat\tau_{\mathrm{DiM}}\,|\,X,T)^2+\mathrm{Var}(\widehat\tau_{\mathrm{DiM}}\,|\,X,T) = (\beta^\top\Delta_X)^2 + \sigma^2\!\left(\tfrac{1}{n_T}+\tfrac{1}{n_C}\right).
\end{align*}

\paragraph{Relating $(\beta^\top\Delta_X)^2$ to $M$} 
Now, how can we bound or approximate $(\beta^\top\Delta_X)^2$ using only the \textit{realized} multivariate distance $M$? Assume columns of $X$ are standardized and pairwise independent (in practice, whiten $X$ so that $\Sigma_X=I_d$). Then $M=\Delta_X^\top \Delta_X = \sum_j \mathrm{SMD}_j^2$, recalling that SMD$_j$ indicates the \underline{s}tandardized \underline{m}ean \underline{d}ifference (SMD) for column/covariate $j$.

Two consequences follow.
\begin{enumerate}[leftmargin=*,nosep]
\item (\emph{Bound}) Cauchy–Schwarz: $(\beta^\top\Delta_X)^2 \le \|\beta\|_2^2 \,\;\; \|\Delta_X\|_2^2 = \sigma_{\text{Prog}}^2\, M$, where $\|\Delta_X\|_2^2 = M$ by definition and since after standardization/whitening, it follows that $\operatorname{Cov}(X_i)=I_d$, and therefore
\[
\|\beta\|_2^2
= \beta^\top \beta
= \beta^\top \operatorname{Cov}(X_i)\beta
= \operatorname{Var}(\beta^\top X_i)
\equiv \sigma_{\text{Prog}}^2.
\]
The Cauchy-Schwarz bound represents the worst-case prognostic direction ($\beta$ perfectly aligned with the realized imbalance vector $\Delta_X$) and is therefore highly conservative in high dimensions. 

\vspace{0.25cm}

The typical-orientation expression described next is the expected squared bias term when the prognostic direction $\beta$ is fixed but the imbalance direction is effectively random, which is the quantity of primary relevance in high-dimensional settings.
\item (\emph{Typical orientation}) Assuming \(\Delta_X\) is approximately multivariate Normal with covariance \((1/n_T+1/n_C)I_d\) (e.g., by a CLT argument), which makes it spherically symmetric, we see that, conditional on $\|\Delta_X\|_2^2 = M$, the direction $\Delta_X/\|\Delta_X\|_2$ is uniform on the unit sphere. Let $u \equiv \beta/\|\beta\|_2$. Then
\begin{align*} 
(\beta^\top \Delta_X)^2 &= \Big(\|\beta\|_2 \big(u^\top \Delta_X\big)\Big)^2 
\\ &= \|\beta\|_2^2 \;\; \big(u^\top \Delta_X\big)^2 
\\ &= \|\beta\|_2^2 \;\; \|\Delta_X\|_2^2 \;\; \Big(u^\top \tfrac{\Delta_X}{\|\Delta_X\|_2}\Big)^2,
\end{align*}
so 
\[
\mathbb{E}\big[(\beta^\top\Delta_X)^2 \,\big|\, \|\Delta_X\|_2^2 = M\big]
= \|\beta\|_2^2 \, M \; \cdot \;
\mathbb{E}\Big[\big(u^\top  \tfrac{\Delta_X}{\|\Delta_X\|_2}\big)^2 \,\Big|\, \|\Delta_X\|_2^2 = M\Big]
= \|\beta\|_2^2 \; M \cdot \frac{1}{d}
= \frac{\sigma_{\text{Prog}}^2}{d}\, M, 
\]
where the first equality holds from: 
\begin{itemize}
    \item[] The decomposition $\beta=\|\beta\|_2u$ and $\Delta_X=\|\Delta_X\|_2\big(\Delta_X/\|\Delta_X\|_2\big)$, so
$(\beta^\top\Delta_X)^2=\|\beta\|_2^2\,\|\Delta_X\|_2^2\,\big(u^\top(\Delta_X/\|\Delta_X\|_2)\big)^2$; conditioning on $\|\Delta_X\|_2^2=M$ fixes the factor $\|\Delta_X\|_2^2$ at $M$.
\end{itemize}
The second equality follows from rotational symmetry of the sphere: 
\begin{itemize}
\item[] If $V$ is uniform on the unit sphere and $w$ is any fixed unit vector, then $\mathbb{E}[(w^\top V)^2]=1/d$ \citep{vershynin2025high}: the typical squared cosine between $w$ and $V$ is $1/d$; intuitively, a random direction in $\mathbb{R}^d$ spends about a $1/d$ fraction of its squared length in any fixed direction, so the average squared alignment with $u$ is $1/d$.\footnote{
More specifically, if $V$ is uniform on $S^{d-1}$ then by symmetry $\mathbb{E}[VV^\top]=c\,I_d$ for some $c$, and taking traces gives
$\mathbb{E}[\|V\|_2^2](=1)=\mathrm{tr}(\mathbb{E}[VV^\top])=cd$; solving for $c$ yields $\mathbb{E}[VV^\top]=I_d/d$ and hence
$\mathbb{E}[(w^\top V)^2]=w^\top \mathbb{E}[VV^\top] w=1/d$ for any unit vector, $w$.
} Substitution into the conditional MSE yields the realized RMSE approximation.
\end{itemize}

\end{enumerate}

\paragraph{Remark (general $\Sigma_X$)} If $\Sigma_X\succ 0$, let $\widetilde X = X\Sigma_X^{-1/2}$, $\widetilde\beta=\Sigma_X^{1/2}\beta$, and $\widetilde\Delta=\bar{\widetilde X}_T-\bar{\widetilde X}_C = \Sigma_X^{-1/2}\Delta_X$. Then $M=\Delta_X^\top\Sigma_X^{-1}\Delta_X = \|\widetilde\Delta\|_2^2$ and $\sigma_{\text{Prog}}^2=\|\widetilde\beta\|_2^2$, allowing application of results from the standardized/whitened case.

\subsection*{B. Ex-ante MSE under complete randomization}

\noindent Under complete randomization with standardized covariates $X$ (so that $\mathbb{E}[X_i]=0$ and $\operatorname{Cov}(X_i)=I_d$), the treated and control sample means are
\[
\bar X_T \equiv \frac{1}{n_T}\sum_{i:T_i=1} X_i,
\qquad
\bar X_C \equiv \frac{1}{n_C}\sum_{i:T_i=0} X_i,
\]
and the covariate mean difference is $\Delta_X \equiv \bar X_T - \bar X_C$. By symmetry of the randomization,
\[
\mathbb{E}[\bar X_T] = \mathbb{E}[\bar X_C] = 0
\quad\Rightarrow\quad
\mathbb{E}[\Delta_X] = 0.
\]

\noindent Treating units as independent draws from a super-population with $\operatorname{Cov}(X_i)=I_d$, the variance of each sample mean is
\[
\operatorname{Var}(\bar X_T) = \frac{1}{n_T} I_d,
\qquad
\operatorname{Var}(\bar X_C) = \frac{1}{n_C} I_d,
\]
and the treated and control groups are independent under the i.i.d. super-population model. Hence, 
\[
\operatorname{Var}(\Delta_X)
= \operatorname{Var}(\bar X_T - \bar X_C)
= \operatorname{Var}(\bar X_T) + \operatorname{Var}(\bar X_C)
= \Big(\tfrac{1}{n_T} + \tfrac{1}{n_C}\Big) I_d.
\]

\noindent Now consider the scalar random variable $\beta^\top \Delta_X$. Using $\mathbb{E}[\Delta_X]=0$,
\[
\mathbb{E}[\beta^\top \Delta_X] = \beta^\top \mathbb{E}[\Delta_X] = 0,
\]
so $\mathbb{E}\big[(\beta^\top \Delta_X)^2\big] = \operatorname{Var}(\beta^\top \Delta_X)$. By the bilinearity of variance for linear forms,
\[
\operatorname{Var}(\beta^\top \Delta_X)
= \beta^\top \operatorname{Var}(\Delta_X)\beta
= \Big(\tfrac{1}{n_T} + \tfrac{1}{n_C}\Big)\beta^\top I_d \beta
= \Big(\tfrac{1}{n_T} + \tfrac{1}{n_C}\Big)\|\beta\|_2^2.
\]
Recall that, under standardized $X$, the prognostic variance is
\[
\sigma_{\text{Prog}}^2 \equiv \operatorname{Var}(\beta^\top X_i)
= \beta^\top \operatorname{Cov}(X_i)\beta
= \beta^\top I_d \beta
= \|\beta\|_2^2,
\]
so
\[
\mathbb{E}\big[(\beta^\top \Delta_X)^2\big]
= \Big(\tfrac{1}{n_T} + \tfrac{1}{n_C}\Big)\sigma_{\text{Prog}}^2.
\]
\noindent From the conditional decomposition (above):
\[
\operatorname{MSE}(\widehat\tau_{\mathrm{DiM}} \mid X,T)
= (\beta^\top \Delta_X)^2
+ \sigma^2\Big(\tfrac{1}{n_T} + \tfrac{1}{n_C}\Big),
\]
taking expectations over the randomization yields
\[
\mathbb{E}\!\left[\operatorname{MSE}(\widehat\tau_{\mathrm{DiM}})\right]
= \mathbb{E}\big[(\beta^\top \Delta_X)^2\big]
+ \sigma^2\Big(\tfrac{1}{n_T} + \tfrac{1}{n_C}\Big)
= \Big(\tfrac{1}{n_T} + \tfrac{1}{n_C}\Big)\left(\sigma^2 + \sigma_{\text{Prog}}^2\right).
\]
We might, however, want to consider how the MSE varies with the rerandomization acceptance rule, which we examine next. 

\subsection*{C. Effect of a Mahalanobis acceptance rule}

\noindent \paragraph{Strategy} Our goal is to understand how a Mahalanobis acceptance rule,
\[
\text{accept} \iff M=\Delta_X^\top\Delta_X \le a,
\]
alters the distribution of the treated--control mean difference $\Delta_X$ and therefore the ex-ante MSE of $\widehat\tau_{\mathrm{DiM}}$.
Under the Gaussian (CLT) approximation with whitened covariates, $\Delta_X$ is approximately isotropic, so conditioning on $M\le a$ is geometrically just truncating an isotropic Gaussian to the Euclidean ball of radius $\sqrt{a}$.
The key observation is that the event $\{M\le a\}$ depends only on the \emph{radius} of $\Delta_X$ and not its \emph{direction}; therefore, the conditional direction remains uniform on the sphere, and the conditional covariance must shrink by a scalar factor found below, yielding the shrunken MSE under rerandomization.

\noindent \noindent \paragraph{Details} Because under the CLT,  $\Delta_X \sim \mathcal{N}\!\left(0,\left(\tfrac{1}{n_T}+\tfrac{1}{n_C}\right)I_d\right)$ is approximately spherically symmetric under the assumptions outlined above, we can write it in spherical coordinates as
\[
\Delta_X = \rho\, U,
\]
where
\begin{itemize}[leftmargin=*,nosep]
\item $\rho \equiv \|\Delta_X\|_2 \ge 0$ is the random radius;
\item $U \equiv \Delta_X / \|\Delta_X\|_2$ is the random direction, which is uniform on the unit sphere $S^{d-1}$.
\end{itemize}
Because the squared norm of a $d$–dimensional Gaussian vector has a $\chi^2_d$ distribution,
\begin{equation}\label{ref:Chi2}
\rho^2 = \|\Delta_X\|_2^2 \sim \Big(\tfrac{1}{n_T}+\tfrac{1}{n_C}\Big)\chi^2_d,
\end{equation}
with $\rho$ independent of the random direction, $U$. The acceptance event \(\{M\le a\}\) is equivalent to \(\{\chi^2_d \le c\}\) with \(c \equiv a / (\tfrac{1}{n_T}+\tfrac{1}{n_C})\). Equipped with this definition of acceptance, we can now evaluate the conditional expectation of the imbalance outer product, which we need because it allows us to compute the expected squared prognostic bias, $(\beta^\top\Delta_X)^2$, via the quadratic form $\beta^\top \mathbb{E}[\Delta_X\Delta_X^\top]\beta$. Using $\mathbb{E}[UU^\top]=I_d/d$ and the independence of $\rho$ and $U$,
\[
\mathbb{E}\big[\Delta_X\Delta_X^\top \,\big|\, M\le a\big]
= \mathbb{E}\big[\rho^2 UU^\top \,\big|\, \rho^2\le a\big] = \mathbb{E}\big[\rho^2 \,\big|\, \rho^2\le a\big]\cdot \mathbb{E}[UU^\top] = \mathbb{E}\big[\rho^2 \,\big|\, \rho^2\le a\big] \cdot I_d / d. 
\]
Rerandomization shrinks the complete-randomization covariance
$\big(\tfrac{1}{n_T}+\tfrac{1}{n_C}\big)I_d$
by a \emph{single scalar} determined by the truncated radial moment
$\mathbb{E}[\rho^2\mid \rho^2\le a]$ (equivalently, by the truncated $\chi^2$ moment). In particular, all directions shrink equally because the acceptance event depends only on the radius and does not privilege any coordinate direction. 

To evaluate the scalar $\mathbb{E}\!\left[\rho^2 \,\middle|\, \rho^2\le a\right]$, use the $\chi^2$ representation in Eq. \ref{ref:Chi2}. Let $Z\sim\chi^2_d$ so that
\[
\rho^2 \;=\; \Big(\tfrac{1}{n_T}+\tfrac{1}{n_C}\Big)Z.
\]
Because $\{M\le a\}\iff\{\rho^2\le a\}$ and $\rho^2$ is a rescaling of $Z$, the acceptance event is equivalently $\{Z\le c\}$ with
\[
c \;\equiv\; \frac{a}{\tfrac{1}{n_T}+\tfrac{1}{n_C}}.
\]
Therefore,
\[
\mathbb{E}\!\left[\rho^2 \,\middle|\, M\le a\right]
=
\Big(\tfrac{1}{n_T}+\tfrac{1}{n_C}\Big)\,
\mathbb{E}\!\left[\chi^2_d \,\middle|\, \chi^2_d\le c\right].
\]
It is convenient to summarize the effect of truncation by the dimension-normalized shrinkage factor
\[
v_a(d)
\;\equiv\;
\frac{\mathbb{E}\!\left[\chi^2_d \,\middle|\, \chi^2_d \le c\right]}{d}
\in (0,1),
\qquad\text{so that}\qquad
\mathbb{E}\!\left[\chi^2_d \,\middle|\, \chi^2_d \le c\right]=d\,v_a(d).
\]
A standard identity for truncated $\chi^2$ moments (see footnote) gives the explicit form\footnote{
Let $X\sim\chi^2_d$ with density
$f_d(x)=\frac{1}{2^{d/2}\Gamma(d/2)}x^{d/2-1}e^{-x/2}$ for $x>0$. Then
\[
\mathbb{E}[X\,\mathbf{1}\{X\le c\}]
=\int_0^c x f_d(x)\,dx
=\frac{1}{2^{d/2}\Gamma(d/2)}\int_0^c x^{d/2}e^{-x/2}\,dx.
\]
Using $\Gamma(d/2+1)=(d/2)\Gamma(d/2)$, the right-hand side can be rewritten as
\[
d\int_0^c f_{d+2}(x)\,dx
= d\,\Pr(\chi^2_{d+2}\le c).
\]
Dividing by $\Pr(\chi^2_d\le c)$ yields
$\mathbb{E}[X\mid X\le c]= d\,\Pr(\chi^2_{d+2}\le c)/\Pr(\chi^2_d\le c)$ and hence
$v_a(d)=\Pr(\chi^2_{d+2}\le c)/\Pr(\chi^2_d\le c)$, as in \citet{MorganRubin2012}.
}
\[
v_a(d)
\;=\;
\frac{\Pr(\chi^2_{d+2}\le c)}{\Pr(\chi^2_d\le c)}.
\]
Substituting this into the previously derived expression
$\mathbb{E}\!\left[\Delta_X\Delta_X^\top \mid M\le a\right]
=\mathbb{E}\!\left[\rho^2 \mid \rho^2\le a\right]\cdot I_d/d$
yields the isotropically shrunken second moment:
\[
\mathbb{E}\!\left[\Delta_X\Delta_X^\top \,\middle|\, M\le a\right]
=
\Big(\tfrac{1}{n_T}+\tfrac{1}{n_C}\Big)v_a(d)\,I_d.
\]
Moreover, $\mathbb{E}[\Delta_X\mid M\le a]=0$ by symmetry (the event $\{M\le a\}$ depends only on $\|\Delta_X\|_2$), so this is also the conditional covariance $\operatorname{Var}(\Delta_X\mid M\le a)$.

Consequently, the expected squared prognostic bias among accepted assignments is
\[
\mathbb{E}\!\left[(\beta^\top\Delta_X)^2 \,\middle|\, M\le a\right]
=
\beta^\top \mathbb{E}\!\left[\Delta_X\Delta_X^\top \,\middle|\, M\le a\right]\beta
=
\Big(\tfrac{1}{n_T}+\tfrac{1}{n_C}\Big)v_a(d)\,\beta^\top\beta
=
\Big(\tfrac{1}{n_T}+\tfrac{1}{n_C}\Big)v_a(d)\,\sigma_{\text{Prog}}^2,
\]
where $\beta^\top\beta=\sigma_{\text{Prog}}^2$ under the standardized/whitened normalization.
Adding the residual-noise variance term from Section~A yields the ex-ante MSE conditional on acceptance:
\[
\mathbb{E}\!\left[\mathrm{MSE}(\widehat\tau_{\mathrm{DiM}})\,\middle|\, M\le a\right]
=
\Big(\tfrac{1}{n_T}+\tfrac{1}{n_C}\Big)\left(\sigma^2 + v_a(d)\,\sigma_{\text{Prog}}^2\right).
\]
Finally, the expected imbalance distance among accepted assignments follows immediately by taking a trace (since
$M=\Delta_X^\top\Delta_X=\mathrm{tr}(\Delta_X\Delta_X^\top)$ and $\mathbb{E}[\mathrm{tr}(A)]=\mathrm{tr}(\mathbb{E}[A])$):
\[
\mathbb{E}\!\left[M \,\middle|\, M\le a\right]
=
\mathbb{E}\!\left[\Delta_X^\top\Delta_X \,\middle|\, M\le a\right]
=
\mathrm{tr}\,\mathbb{E}\!\left[\Delta_X\Delta_X^\top \,\middle|\, M\le a\right]
=
\Big(\tfrac{1}{n_T}+\tfrac{1}{n_C}\Big)\, d \, v_a(d).
\]

\subsection*{D. Design-stage inversion for choosing a threshold}

\noindent For a target two-sided test with size $\alpha$ and power, $1-\tilde{B}$, a Normal approximation\footnote{
Assuming \(\hat{\tau} \sim \mathcal{N}(\tau, \operatorname{RMSE}_{\hat{\tau}}^2)\), the standardized statistic \(Z = \frac{\hat{\tau} - \tau}{\operatorname{RMSE}_{\hat{\tau}}} \sim \mathcal{N}(0,1)\). Under \(H_0: \tau = 0\), \(Z = \frac{\hat{\tau}}{\operatorname{RMSE}_{\hat{\tau}}} \sim \mathcal{N}(0,1)\). Under \(H_1: \tau \neq 0\), \(\frac{\hat{\tau}}{\operatorname{RMSE}_{\hat{\tau}}} = Z + \frac{\tau}{\operatorname{RMSE}_{\hat{\tau}}} \sim \mathcal{N}\left( \frac{\tau}{\operatorname{RMSE}_{\hat{\tau}}}, 1 \right)\).
}
implies $|\tau|/\operatorname{RMSE}\gtrsim z_{1-\alpha/2}+z_{1-\tilde{B}}$, meaning that the signal-to-noise ratio $|\tau|/\operatorname{RMSE}$ must be at least $z_{1-\alpha/2}+z_{1-\tilde{B}}$ to achieve size $\alpha$ and power $1-\tilde{B}$ \citep{rosner2006fundamentals}.\footnote{For a two-sided $z$-test that rejects when $|\hat{\tau}|/\operatorname{RMSE} > z_{1-\alpha/2}$, the Normal approximation gives
\[
\Pr(\text{reject}\mid \tau) \;=\; \Pr\!\left(|Z+\tfrac{\tau}{\operatorname{RMSE}}|> z_{1-\alpha/2}\right),
\quad Z\sim\mathcal{N}(0,1).
\]
A conservative sufficient condition for achieving power $1-\tilde{B}$ is
\[
\frac{|\tau|}{\operatorname{RMSE}} \;\ge\; z_{1-\alpha/2}+z_{1-
\tilde{B}
}.
\]
} Using the conditional-on-acceptance MSE,
\[
\operatorname{RMSE}^2(a)
=\Big(\tfrac{1}{n_T}+\tfrac{1}{n_C}\Big)\big(\sigma^2 + v_a(d)\sigma_{\text{Prog}}^2\big)
\le \frac{\tau^2}{\big(z_{1-\alpha/2}+z_{1-\tilde{B}}\big)^2},
\]
where the MSE we are willing to tolerate after rerandomization cannot be larger than the square of the effect size divided by the usual Normal critical value sum for the desired power. Finally, one solves for the largest $a$ (or, equivalently, acceptance probability $q$) such that $v_a(d)$ meets the inequality. Because $v_a(d)$ is strictly decreasing in stringency, a scalar line search suffices. In \pkg{fastrerandomize}, this calculation is wrapped by \code{diagnose\_rerandomization()} to return a recommended \code{randomization\_accept\_prob} given user-specified contextual factors. 

\section{Sample code snippets}
\label{subsec:snippets}

\noindent\textbf{Installation and backend setup}

{\color{gray}
\begin{verbatim}
# install.packages("devtools")
devtools::install_github("cjerzak/fastrerandomize-software/fastrerandomize")
library(fastrerandomize)

# Create/update a conda backend for JAX - Done once upon installation
# build_backend(conda_env = "fastrerandomize") 
\end{verbatim}
}

\noindent\textbf{Generate assignments via Monte Carlo}

{\color{gray}
\begin{verbatim}
set.seed(456)
n_units   <- 200
n_treated <- 100
X <- matrix(rnorm(n_units * 50), nrow = n_units)  # 50 covariates

rand_mc <- generate_randomizations(
  n_units                   = n_units,
  n_treated                 = n_treated,
  X                         = X,
  randomization_type        = "monte_carlo",
  randomization_accept_prob = 0.01,   # keep top 1%
  max_draws                 = 2e5,
  batch_size                = 1e4,
  approximate_inv           = TRUE
)
\end{verbatim}
}

\noindent\textbf{Design-respecting randomization test}

{\color{gray}
\begin{verbatim}
obsW <- rand_mc$randomizations[1, ] 
beta <- rnorm(ncol(X))
tau  <- 1
obsY <- as.numeric(X %*% beta + tau * obsW + rnorm(n_units, 0, 0.5))

test_out <- randomization_test(
  obsW                     = obsW,
  obsY                     = obsY,
  candidate_randomizations = rand_mc$randomizations,
  findFI                   = TRUE
)
\end{verbatim}
}

\end{document}